\documentclass[namedreferences]{solarphysics}

\usepackage[hyperref,optionalrh]{spr-sola-addons} % For Solar Physics 
\usepackage{graphicx, geometry}        % For eps figures, newer & more powerfull

\usepackage{color}           % For color text: \color command
\usepackage{breakurl}        % For breaking URLs easily trough lines

\usepackage[T1]{fontenc}
\usepackage [latin1] {inputenc}

\usepackage{graphicx}
\chardef\us=`\_

\begin{document}

\begin{article}
\begin{opening}

\title{Kinematic study of radio-loud CMEs associated with solar flares and DH type II radio emissions during solar cycles 23 and 24}

\author[addressref=aff1,corref,email={kalai2230@gmail.com}]{\fnm{P}~\lnm{Pappa Kalaivani}\orcid{0000-0002-7364-0768}}
\author[addressref={aff2,aff3}, email={prakash18941@gmail.com}]{\fnm{O}~\lnm{Prakash}\orcid{0000-0002-3312-0629}}
\author[addressref={aff4},email={ashanmugaraju@gmail.com}]{\fnm{A}~\lnm{Shanmugaraju}\orcid{0000-0002-2243-960X}}
%\author[addressref={aff3},email={lfeng@pmo.ac.cn}]{\fnm{}~\lnm{Li Feng}\orcid{0000-0003-3636-2962}}
%\author[addressref={aff3},email={leilu@pmo.ac.cn}]{\fnm{}~\lnm{Lei Lu}\orcid{0000-0002-3032-6066}}
%\author[addressref={aff3},email={wqgan@pmo.ac.cn}]{\fnm{}~\lnm{Weiqun Gan}}%\orcid{0000-0002-2243-960X}}
\author[addressref={aff5}, email={michalek@oa.uj.edu.pl}]{\fnm{G}~\lnm{Michalek}}%\orcid{0000-0002-2243-960X}}
\author[addressref={aff6}, email={selvarani.ganesan@gmmail.com}]{\fnm{G}~\lnm{Selvarani}}%\orcid{0000-0002-2243-960X}}

%\author{\inits{}\fnm{}~\lnm{}\orcid{}}
%   NOTE:  Just one corresponding author [corref]
%   \institute{$^{1}$ First affiliation
%                     email: \url{e.mail-a} email: \url{e.mail-b}\\ 
%              $^{2}$ Second affiliation
%                     email: \url{e.mail-c} \\
%             \textit{}
\address[id=aff1]{Department of Physics, Ultra College of Engineering and Technology for Women, Ultra Nagar, Madurai - 625 104, Tamil Nadu, India}
\address[id=aff2]{Department of Physics, Sethu Institute of Technology, Pulloor, Kariapatti,Viruthunagar, Tamil Nadu -- 626 115, India}
\address[id=aff4]{Department of Physics, Arul Anandar College, Karumathur, Madurai - 625 514, India}
\address[id=aff3]{Key Laboratory of Dark Matter and Space Astronomy, Purple Mountain Observatory, Chinese Academy of Sciences, Nanjing -- 210008, Jiangsu, China}
\address[id=aff5]{Astronomical Observatory of Jagiellonian University, Cracow, Poland}
\address[id=aff5]{Department of Physics, Sri Meenakshi Govt. Arts College for Women, Madurai, India}

%\author[addressref={aff1,aff2,aff3},email={e-mail.a@mail.com}]{\inits{F.N.}\fnm{First~Names}~\lnm{Last~Name~Author-a}\orcid{123-456-7890}}
%\author[addressref=aff1,email={e-mail.b@mail.com}]{\inits{F.}\fnm{First~Names}~\lnm{Last~Name~Author-b}}
%\author[addressref=aff2,corref,email={e-mail.c@mail.com}]{\inits{F.}\fnm{First~Names}~\lnm{Last~Name~Author-c}\orcid{987-654-3210}}
%\author[addressref=aff3]{\inits{T.}\fnm{First~Names}~\lnm{Last~Name~Author-d}}
%\author{\inits{}\fnm{}~\lnm{}\orcid{}}
%   NOTE:  Just one corresponding author [corref]
%   \institute{$^{1}$ First affiliation
%                     email: \url{e.mail-a} email: \url{e.mail-b}\\ 
%              $^{2}$ Second affiliation
%                     email: \url{e.mail-c} \\
%             \textit{}
%\address[id=aff1]{First very very very very very very very very very
 %very very very very very very very very very very very very very very very very very very long affiliation and address}
%\address[id=aff2]{Institution, City, State, Country}
%\address[id=aff3]{Third affiliation and address}

\runningauthor{Pappa Kalaivani et al. 2022}
%\runningtitle{\textit{Solar Physics} Example Article}

\begin{abstract}
We have statistically analyzed 379 radio-loud (RL) CMEs and their associated flares during the period 1996 - 2019 covering both solar cycles (SC) 23 and 24. We classified them into two sets of populations based on the observation period: i) 235 events belong to SC 23 (August 1996 - December 2008) and ii) 144 events belong to SC 24 (January 2009 - December 2019).For both cycles, the mean sky-plane speed, projection corrected speed (space speed), and initial acceleration of RL CMEs are found to be similar. Moreover, the average residual acceleration of RL CMEs in SC 24 (--17.39 $\pm$ 43.51 m s$^{-2}$) is two times lower than that of the RL CMEs in SC 23 (--8.29 $\pm$ 36.23 m s$^{-2}$), which means that deceleration of RL CMEs in SC 24 is twice as fast as in SC 23. RL CMEs of SC 23 (1443 $\pm$ 504 km s$^{-1}$; 13.82 $\pm$ 7.40 \emph{R}$_{\circledcirc}$) reach their peak speed at higher altitudes than RL CMEs of SC 24 (1920 $\pm$ 649 km s$^{-1}$; 12.51 $\pm$ 7.41 \emph{R}$_{\circledcirc}$).We also observed that the mean apparent widths of RL CMEs in SC 23 are less than in SC 24which is statistically significant. SC 23 has a lower average CME nose height (3.85 \emph{R}$_{\circledcirc}$) at the start time of DH type II bursts than that of SC 24 (3.46 \emph{R}$_{\circledcirc}$). The starting frequencies of DH type II bursts associated with RL CMEs for SC 24 are significantly larger (formed at lower heights) than that of SC 23.  We found that there is a good correlation between the drift rates and the mid-frequencies of DH type II radio bursts for both the solar cycles (\emph{R} = 0.80, $\epsilon$ = 1.53). Most of the RL CMEs kinematics and their associated solar flare properties are found similar for SC 23 and SC 24. The annual variations for general CMEs are well consistent with mean sunspot number but small variations in halo and RL CMEs are observed. We concluded that the reduced total pressure in the heliosphere for SC 24 enables RL CMEs to expand wider and decelerate faster, resulting in DH type II radio emissions at lower heights than SC 23.
  
\end{abstract}
\end{opening}
%-------------------------------------------------

\section{Introduction}
     \label{S-Introduction} 
     
    % \section{Introduction}
     %\label{S-Introduction} 

CMEs (coronal mass ejections) are important components that can alter geo-space weather conditions. CMEs are highly magnetised plasma eruptions from the Sun that propagate into the solar corona and interplanetary medium (IP) with an average mass of 10$^{16}$ kg and an extreme energy of 10$^{32}$ ergs (Vourlidas et al., 2010; Gopalswamy, 2016). The Lorentz force, aerodynamic drag, and gravitational forces have all been linked to CME dynamics. The CME emerges from the rest with a large initial acceleration in the early stages. Aerodynamic drag affects the relative velocities of CMEs and solar wind, which can lead CMEs to accelerate or decelerate (Vrsnak et al., 2004; Vrsnak, 2006).The corona and interplanetary (IP) medium have a significant impact on CME propagation, owing to interactions between CMEs and ambient turbulence and hot magnetic plasma (Cargill et al., 1996; Cargill, 2004). Whether a CME's initial acceleration is impulsive or gradual depends on whether the CMEs originate in the active or quiet areas of the Sun (Vrsnak, 2001; Cliver et al.,2004; Zhang and Dere, 2006).It is generally understood that CMEs linked with prominences are slower, while CMEs related with solar flares are more impulsive (Sheeley et al., 1999). \indent The physical relationship between solar flares and CMEs is still a subject of debate (Webb and Howard, 2012).\\[5pt]
White-light coronagraphic observations from onboard and ground-based devices have made the most significant contributions to the understanding of CMEs. The Large Angle and Spectrometric Coronagraph (LASCO, Brueckner et al., 1995) on board the Solar and Heliospheric Observatory (SOHO) mission (Domingo, Fleck, and Poland, 1995) has been one of the most important instruments for studying CMEs for the past two decades. The Coordinate Data Analysis Workshop (CDAW) data base contains a monthly CMEs list in which CMEs are spotted from synaptic images and fundamental properties of CMEs are manually determined from height-time (h-t) measurements before being included in SOHO/LASCO CME online catalogue1 (Yashiro et al., 2004; Gopalswamy et al., 2009a). Based on Hundhausen et al. (1984) description of CMEs, many authors have stated that human observation is the best way for identifying CMEs (Yashiro, Michalek, and Gopalswamy, 2008; Hess, and Colaninno, 2017; Vourlidas et al., 2017; Webb et al., 2017). The SOHO/LASCO instrument has three coronagraphs C1, C2, and C3, that cover the observation ranges of 1.1 - 3 \emph{R}$_{\circledcirc}$, 2 - 6.5 \emph{R}$_{\circledcirc}$, and 3.8 - 30 \emph{R}$_{\circledcirc}$, respectively. Unfortunately, after 2.5 years of observation, the innermost C1 coronagraph perished. The SOHO/LASCO coronagraphs, on the other hand, can track CMEs up to a maximum heliocentric distance of 32 \emph{R}$_{\circledcirc}$. Based on the leading edge (LE) observations, the h-t images are manually measured following the fastest feature of a particular CME. Except for the two slightly longer data gaps in 1998 and 1999, CME observations have been nearly continuous across the whole span of two Solar Cycles (hereinafter SC 23 and 24). The CDAW's SOHO/LASCO catalogue recorded 30228 CMEs to May 2020 and identified the key features of CMEs. The halo CMEs receives special attention among all CMEs. In sky-plane projection, the halo CMEs seem to surround the occulting disc of the observing white-light coronagraphs (Howard et al., 1982). Halo CMEs are generally faster and wider than normal CMEs, and they are also associated with stronger (M and X class) solar flares (Gopalswamy, Yashiro, and Akiyama, 2007). Nearly 3\% of overall populations of CMEs are halo (Lamy et al., 2019). The geo-effectiveness of halo CMEs is frequently higher than that of regular CMEs. (Lara et al. 2006). Furthermore, disc halo CMEs have been found in the literature to be the most likely to cause major geomagnetic storms (Kim et al. 2005; Michalek et al. 2006; Yermolaev and Yermolaev, 2006; Prakash et al., 2014; Scolini, Messerotti, and Poedts, 2018).\\[5pt]
Another type of CMEs, referred to as radio-loud (RL) CMEs, is more fascinating to study and forecast space weather conditions (Gopalswamy et al., 2012b). These CMEs are associated to type II radio bursts and, as a result, to CME-driven shocks. The radio emission in the solar corona and IP medium provides important evidence to CME-driven shocks acceleration and propagation. Gopalswamy et al. (2005) found that the mean speed and width of type II associated CMEs are 1115 km sc and 139$^{o}$, respectively, and 42.5\% of CMEs are halo. CMEs associated with solar energetic particles (SEPs)and type II bursts are wider and faster than the general CME (Gopalswamy et al., 2008).Gopalswamy et al. (2012a) developed a comprehensive catalog that includes the features of parent solar eruptions (CME and solar flares) and their associated type II radio bursts. They used type II radio bursts recorded in the decameter-hectometric (DH) domain as the primary data obtained by the Radio and Plasma Wave Experiment (WAVES, Bougeret et al., 1995) on board the Wind spacecraft (hereinafter Wind/WAVES) only prior to October 2006 and thereafter by both Wind/WAVES and the Solar Terrestrial Relation Observatory (Kaiser et al., 2008) on board Sun Earth Connection Coronal and Heliospheric Investigation (Howard et al., 2008) or STEREO/WAVES. Because CME-driven shocks accelerate solar energetic particles and can produce geomagnetic storms, such RL CMEs can affect the Earth magnetosphere, but not all RL CMEs are capable of doing so.\\[5pt]
The main features of space weather events that occurred around the two peaks in the sunspot number (SSN) during SC 24 were compared by Gopalswamy et al. (2015a). They identified that all SEP events were associated to DH type II bursts, which are better indicators of space weather than halo CMEs. They also found that for the first 73 months of each cycle, the mean speed of CMEs associated with strong geomagnetic storms in SC 24 (968 km s$^{-1}$) is higher than in SC 23 (838 km s$^{-1}$). Michalek, Gopalswamy, and Yashiro (2019) investigated the rate of CME detection by SOHO/LASCO during SC 23 and SC 24. They found that in SC 24, the average characteristics of CMEs altered markedly.  CMEs seem to be narrower, slower, ejected at a wider range of latitudes, and slightly faster than those observed in SC 23 on average. Lamy et al. (2019) presented a review article on statistical analysis of CME features based on 23 years of quasi-continuous observations with the SOHO/LASCO coronagraph for SC 23 and SC 24 in another study. For the general population of CMEs, they found that CME activity was higher in SC 24 than in SC 23. To better understand the status of the heliosphere throughout these solar cycles, Gopalswamy, Akiyama, and Yashiro (2020) compared the features of limb CMEs for SC 23 and SC 24. SC 23 has much faster limb halo CMEs (1637 km s$^{-1}$) than SC 24 (1281 km s$^{-1}$), according to the researchers. They also stated that due to the weak heliospheric state, SC 24 CMEs become halos sooner and at lower speeds than SC 23 CMEs. Shanmugaraju et al. (2021) also used the linear correlations method with the SSN to forecast the occurrence rate of the RL and halo CMEs for the SC 25. They reported that SC 24 has a relative occurrence rate of RL CMEs that is 0.52 times lower than SC 23. They also mention that research on RL CMEs is mainly concerned with space-weather impacts\\[5pt]
Because of these differences, we decided to compare the parameters of RL CMEs, solar flares, and DH type II bursts for SC 23 and SC 24. The objective of this paper is to show the varied early kinematics of RL CMEs detected by the SOHO/LASCO coronagraphs for SC 23 and SC 24 as a whole. A full comparative examination of the kinematics of RL CMEs was also attempted. Furthermore, we investigate the fluctuations in these two solar cycles by calculating the SC 24/SC 23 ratios for the various features of solar flares and DH type II bursts. Section 2 describes the data selection and analysis procedures employed in this investigation. Section 3 presents the statistical analysis and results for RL CMEs, solar flares, and DH type II burst characteristics for SC 23 and SC 24, followed by a summary and conclusions in Section 4.

\section{Event selection and analysis}
Total of 518 RL CMEs were identified from April 1997 to September 2017and catalogued online by the CDAW data centre (hence RL CME catalogue: Gopalswamy, Makela, and Yashiro, 2019). It almost covered two cycles and it exclusively used observations from the SOHO/LASCO coronagraphs. This research spans the SC 23 (August 1996 - December 2008) and SC 24 (January 2009 - December 2019) study periods. But, the RL CME data for the beginning and end of solar minimum for SC 23 (8 months) and SC 24 (27 months), respectively, are not reported in the online catalogue. However, the probability of RL CMEs occurring at these minimum periods is comparatively low. Furthermore, 379 RL CMEs are considered with greater than three h-t measurements, and the remaining 139 RL CMEs were excluded from this study to ensure the derived attributes of RL CMEs are reliable. In addition, these 379 RL CMEs are separated into two populations: \emph{i)} 235 RL CMEs are attributed to SC 23; \emph{ii)} 144 RL CMEs are attributed to SC 24.  In addition to the original set, a subset of limb events (source longitude $>$ 60 degree) are also considered for statistical analysis consisting of SC 23 (65 events) and SC 24 (40 events). The sky-plane speed and width of RL CME, eruption location, flare importance (peak flux), starting frequency, end frequency, start and end time of DH type II bursts are all listed in the RL CME catalogue. Other essential RL CME properties (acceleration, mass, and kinetic energy) from the LASCO CMEs catalogue that are not reported in the RL CMEs catalogue have also been included. As we discussed earlier, the space speed (projection corrected speed) for halo CMEs are adopted and for non-halo CMEs are derived from the sky-plane speed and longitude of the source location using the Cone model (Xie et al., 2004). The geometrical correction has done for non-halo CMEs by multiplying the derived sky-plane speed by the factor of, where ? is the angular distance of a CME source from the center. The initial and final speeds of RL CMEs were also calculated using the first and last two h-t observations, respectively. The peak speed and peak speed height of the RL CMEs were also calculated. The mass of a CME is estimated using a larger number of assumptions. As a result, mass and kinetic energy are highly uncertain and are simply representative. The integrated flux of solar flares is also used from the GOES soft X-ray flare data which is not listed in the RL CMEs catalogue. Table 6.1shows the statistical mean (median) and standard deviation ($\sigma$) of the parameters describing RL CME and soft X-ray solar flare features for SC 23 and SC 24 separately. The features of RL CMEs and solar flares are listed in Column 1. For all RL CMEs, the estimated values of the events mean (median) and standard deviation ($\sigma$) are given in columns 2 \& 3 (SC 23 and SC 24), 4 \& 5 (SC 23), and 6 \& 7 (SC 24), respectively. In Column 8, the ratio of SC 24 to SC 23 (SC 24/SC 23) is given. The results of statistical analysis for SC 23 and SC 24 are presented in the next section along with some discussion.

\newgeometry{margin=1.5in}

\begin{table}[htp]
%\begin{threeparttable}
\caption{Mean (median) and standard deviation ($\sigma$) values of different properties of RL CMEs for SC 23 and SC 24.}
\begin{center}
\begin{tabular}{|l|c|c|c|c|c|c|c|}
\hline
Properties&\multicolumn{2}{c}{SC 23 + SC 24}&\multicolumn{2}{c}{SC 23}&\multicolumn{2}{c}{SC 24}&\shortstack{SC 24 /\\ SC 23}\\
\cline{2-3} \cline{4-5} \cline{6-7}
&Mean (Median)&$\sigma$&Mean (Median)&$\sigma$&Mean (Median)&$\sigma$&\\\hline
Sky-plane Speed (km s$^{-1}$) &1149 (1085)&487&1153 (1092)&460&1142 (1064)&531&0.99\\\hline
Space speed (km s$^{-1}$)&1408 (1335)&576&1403 (1312)&521&1417 (1358)&657&1.01\\\hline
Residual Acceleration  (m s$^{-2}$)&-11.75 (-8.10)&39.35&-8.29 (-3.70)&36.23&-17.39 (-9.80)&43.51&2.10\\\hline
Initial acceleration ( km s$^{-2}$)&1.30 (1.00)&1.12&1.34 (1.03)&1.18&1.22 (0.96)&0.99&0.91\\\hline
Initial speed (km s$^{-1}$)&1207 (1146)&615&1164 (1131)&542&1279 (1194)&714&1.10\\\hline
Final Speed (km s$^{-1}$)
&1195 (1121)
&582
&1149 (1112)
&466
&1270 (1155)
&729
&1.11\\\hline
Peak Speed (km s$^{-1}$)
&1624 (1527)
&609
&1443 (1402)
&504
&1920 (1813)
&649
&1.33\\\hline
Peak Speed height (km s$^{-1}$)
&13.32 (10.99)
&7.42
&13.82 (11.77)
&7.40
&12.51 (8.81)
&7.41
&0.90\\\hline
Width (deg)
&276 (360)
&106
&262 (360)
&112
&299 (360)
&91
&1.14\\\hline
%Mass (g)
%&1.12 $\times$10$^{16}$ (8.95$\times$10$^{15}$)
%&1.18$\times$10$^{16}$
%&1.05$\times$10$^{16}$ (8.20$\times$10$^{15}$)
%&1.33$\times$10$^{16}$
%&1.23$\times$10$^{16}$ (1.00$\times$10$^{15}$)
%&9.28$\times$1015
%&1.18\\\hline
%Kinetic energy (erg)
%&1.21$\times$10$^{32}$ (5.05$\times$10$^{31}$)
%&2.96$\times$10$^{32}$
%&1.12$\times$10$^{32}$ (4.85$\times$10$^{31}$)
%&3.24$\times$10$^{32}$
%&1.33$\times$10$^{32}$ (5.55$\times$10$^{31}$)
%&2.53$\times$10$^{32}$
%&1.19\\\hline
DH starting Frequency (kHz)
&10606 (14000)
&4638
&9945 (13000)
&4710
&11670 (14000)
&4327
&1.17\\\hline
DH ending frequency (kHz)
&2151 (1000)
&2741
&2363 (1150)
&2741
&1810 (400)
&2716
&0.77\\\hline
CME nose height (\emph{R}$_{o}$) 
&3.72 (3.44)
&1.22
&3.85 (3.64)
&1.04
&3.46 (3.20)
&1.48
&1.11\\\hline
Latitude (deg)
&-0.70 (1.00)
&16.78
&-0.89 (0.00)
&17.30
&-0.40 (4.00)
&15.99
&0.45\\\hline
Longitude (deg)
&6.12 (12.00)
&51.43
&7.29 (13.50)
&52.19
&4.29 (12.00)
&50.37
&0.59\\\hline
%Peak flux (W m$^{-2}$)
%&8.15$\times$10$^{-5}$ (3.00$\times$10$^{-5}$)
%&1.44$\times$10$^{-4}$
%&8.02$\times$10$^{-5}$ (2.80$\times$10$^{-5}$)
%&1.36$\times$10$^{-4}$
%&8.41$\times$10$^{-5}$ (3.00$\times$10$^{-5}$)
%&1.59$\times$10$^{-4}$
%&1.05\\\hline
Integrated intensity (J m$^{-2}$)
&0.12 (0.05)
&0.19
&0.12 (0.05)
&0.21
&0.11 (0.05)
&0.13
&0.87\\\hline

\end{tabular}
\end{center}
\end{table}%

\restoregeometry

\section{Results}
\subsection{Radio-Loud CME properties during Solar Cycles 23 and 24}

The features of all CMEs are subject to projection effects, according to Gopalswamy et al. (2000). Many researchers have stated that the projection effect is more essential when studying the properties of CMEs (Michalek, Gopalswamy, and Yashiro, 2005; Wen, Maia, and Wang, 2007; Pappa Kalaivani et al., 2010). Because the longitude (L) of the CME source is greater than 60o, so-called limb events have lesser projection effects than disc events (E30 $\ge$ L $\ge$W30). As previously stated, the LASCO coronagraph images can only detect apparent CME speeds and widths. As a result, there will always be uncertainty in the derived parameters of CMEs. The distributions of (a-c) residual acceleration, (d-f) peak speed, and (g-i) peak speed height for RL CMEs are shown in Figure 1.  In the first column of this figure, the residual acceleration for all RL CMEs (a), SC 23 (b) and SC 24 (c) are shown. The reported acceleration in the CMEs catalog is considered to be \lq\lq residual acceleration\rq\rq, because the acceleration due to gravity and propelling forces must have declined significantly, the deceleration due to drag becomes dominant at heights corresponding to the LASCO field of view (FOV). The residual acceleration ranging from --240 m s$^{-2}$ to 166 m s$^{-2}$ with a mean of --11.75 $\pm$ 39.35 m s$^{-2}$. The distributions of the residual accelerations are similar in all cases, as seen in this figure, and the peak bins are at zero mark. Within the LASCO FOV, about 51\% of RL CMEs are decelerated ($<$ --7.5 m s$^{-2}$), 22\% of RL CMEs are accelerated ($>$ 7.5 m s$^{-2}$) and, the remaining 27\% of RL CME are considered constant speed (--7.5 m s$^{-2}$ $>$ a $>$ 7.5 m s$^{-2}$). Even though the mean speeds of both sets of RL CMEs are the same, we found that the RL CMEs in SC 24 are decelerating faster than the RL CMEs in SC 23. The mean residual accelerations for SC 23 and SC 24 are completely different (--8.29 $\pm$ 36.23 m s$^{-2}$ and --17.39 $\pm$ 43.51 m s$^{-2}$, respectively), the difference in the mean is statistically significant (student \emph{t}-test gives P = 2\%, \emph{i.e.,} indicating that there is a probability of 2\% possibility that the true mean values of the two underlying populations are identical). \\[5pt]
If we simply consider the limb events, we have the same difference between the SC 23 and SC 24 in the residual acceleration. It also reveals that the difference in mean residual acceleration for SC 23 (--5.61 $\pm$ 38.79 m s$^{-2}$) and SC 24 (--18.28 $\pm$ 16.18 m s$^{-2}$) is statistically significant (P = 0.0001\%). Prakash et al. (2014) found that the acceleration of the geo-effective RL CMEs and non-geo-effective RL CMEs during the period 1997 -- 2005 (SC 23). They found that the geo-effective RL CMEs (--14.35 m s$^{-2}$) are more decelerated than the non-geo-effective RL CMEs (--6.57 m s$^{-2}$). It is worth noting that while Prakash et al. (2014) found a similar trend in SC 23, the mean acceleration of the RL CMEs (--17.39 m s$^{-2}$) is significantly smaller than the mean acceleration of the geo-effective RL CMEs (--14.35 m s$^{-2}$) in SC 24.RL CMEs in SC 24 are, on average, two times as strongly decelerated than those in SC 23.The residual acceleration yields a very different conclusion. The anomalous behaviour of CMEs in the SC 24 is assumed to be the cause of a sudden large drop in the mean residual acceleration, even though both sky-plane speed and space speed (projection corrected speed) were found to be similar in both cases.\\[5pt]
Michalek, Gopalswamy, and Yashiro (2017) pointed out that the errors in the estimation of velocity and acceleration significantly depend on the number of height-time points measured for a given CME. Recently, Anitha, Michalek, and Yashiro (2020) used five points linear fit technique to obtain smooth profiles of instantaneous speeds versus distance from the Sun. From this profile, they obtained different kinematic properties of a CME (peak speed, residual velocity, initial acceleration etc). We also utilized their technique but for three points linear fit to obtain the peak speeds and peak speed heights of a set of RL CMEs. Within the LASCO FOV, the peak speed height is specified as the height at which the CMEs reached their maximum speed. It is important to note that the time cadence in the CME observations for SC 23 and SC 24 are different (almost 10 min). This difference in the time cadence between SC 23 and SC 24 do affect the estimation of peak speed height with an error of $\pm$1 \emph{R}$_{\circledcirc}$but not the peak speed. Because, when the CMEs reach their peak speed its acceleration is almost zero. Thus the error in the estimation of peak speed is minimum. The distributions of peak speed is shown in Figure 1 (second column), separately for all RL CMEs (d), SC 23 (e), and SC 24 (f). Fort both SC, RL CMEs have a wide range of peak speeds, ranging from 393 km s$^{-1}$ to 4824 km s$^{-1}$, with a mean of 1624 $\pm$ 609 km s$^{-1}$. Interestingly, 92\% of the peak speed of all RL CMEs (both SC 23 and SC 24) is larger than 900 km s$^{-1}$. However, the difference between the mean peak speeds of RL CMEs for SC 23 (1443 $\pm$ 504 km s$^{-1}$) and SC 24 (1920 $\pm$ 649 km s$^{-1}$) is much smaller. The same trend can be also seen in limb events. The difference in mean peak speed of limb RL CMEs between SC 23 (1594 $\pm$  467 km s$^{-1}$) and SC 24 (2159 $\pm$  799 km s$^{-1}$) is statistically significant (P$<$ 0.1\%).

\begin{figure} [ht]   %%%%%%%%%%%%%%%%%% FIGURE 2 
\centerline{\includegraphics[width=1.\textwidth,clip=]{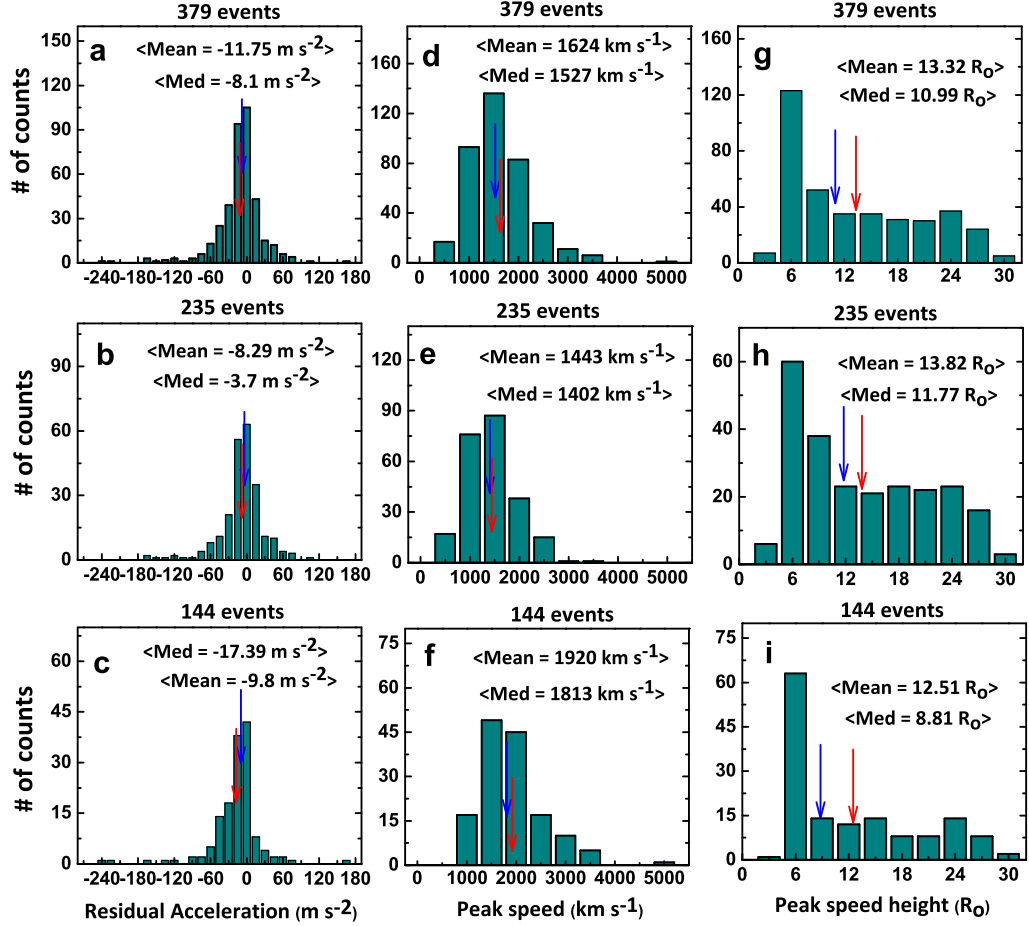}}
\caption{Distributions of (a-c) residual acceleration, (d-f) peak speed, and (g-i) peak speed height for all RL CMEs (top row), SC 23(middle row) and SC 24 (bottom row) events. The mean and median values are marked by red and blue arrows, respectively.}
\end{figure}

Figure 1 (third column) shows the distribution of peak speed height for all RL CMEs (g), SC 23 (h), and SC 24 (i). The mean peak speed height of all RL CMEs is 13.32 $\pm$ 7.42 \emph{R}$_{\circledcirc}$, with a large range of 2.74 \emph{R}$_{\circledcirc}$ to 29.95 \emph{R}$_{\circledcirc}$, as seen in this figure. RL CMEs in SC 23 (13.82 $\pm$ 7.40 \emph{R}$_{\circledcirc}$) reached their peak speed at a higher altitude on average than RL CMEs in SC 24 (12.51 $\pm$ 7.41 \emph{R}$_{\circledcirc}$). The difference between the SC 23 and SC 24 in mean peak speed height is statistically insignificant (P = 9\%).The difference between two sets of events is significantly increased when we consider limb events. The mean difference between SC 23 and SC 24 (15.52 $\pm$ 7.51\emph{R}$_{\circledcirc}$ and 12.61 $\pm$ 7.40 \emph{R}$_{\circledcirc}$, respectively) is statistically significant (P = 5\%), as per student \emph{t}-test. Decelerating CMEs reached their peak speed at lower heights than accelerating CMEs, according to Gopalswamy et al. (2012b).These results indicate that the more decelerating RL CMEs in SC 24 reached their peak speed at a lower altitude than the less decelerating RL CMEs in SC 23. Our conclusion is in line with previous findings for different sets of events (Gopalswamy et al., 2012b; Prakash et al., 2012; 2017). Because of the lowered total pressure (plasma + magnetic) of the interplanetary medium, the different behaviour of the peak speed and their peak speed heights of RL CMEs for SC 24 clearly indicates that CMEs expand (width) rapidly and approaches peak speed at lower heights. Because the internal magnetic field diffused quickly, the propelling force (Lorentz force) reduced quickly.\\[5pt]
  However, the other kinematic parameters of limb RL CMEs (sky-plane speed, space speed, initial acceleration, initial and final speeds) for SC 23 and SC 24 found similar for both sets of SC 23 and SC 24 events. The initial acceleration is estimated using the assumption that CME erupted from the flare reconnection site. The rising phase of a solar flare is considered to be exactly the same as the accelerating phase of a CME (Zhang and Dere, 2006). So, beyond the deceleration due to Sun's gravity, the initial acceleration is primarily owing to propelling force of the CMEs. Because CMEs lift off from rest, initial acceleration in the inner corona is positive. Initial proxy acceleration was calculated using the flare rising time and space speed of RL CMEs in this study. The time difference between the flare peak and flare start of the associated RL CMEs is used to calculate the flare rising time. In addition to that the initial and final speeds of RL CMEs are calculated using the first and last two h-t measurements in the LASCO field of view. The sky-plane speed of all RL CMEs ranges from 291 to 3163 km s$^{-1}$, with an average value of  1149 $\pm$ 487 km s$^{-1}$. Michalek, Gopalswamy, and Yashiro (2019) found that the average speed (391 km s$^{-1}$) of almost 25966 overall populations of CMEs between 1996 and 2015. This speed was over three times lower than that of the mean sky-plane speed of RL CMEs (1149 $\pm$ 487 km s$^{-1}$). Almost 70\% of events have a sky-plane speed of $>$ 900 km s$^{-1}$. When we consider only the limb events, there is significant difference between sky-plane speed (space speed) for SC 23 and SC 24: 1307 $\pm$ 457 km s$^{-1}$ (1344 $\pm$ 448 km s$^{-1}$) and 1296 $\pm$ 641 km s$^{-1}$ (1297 $\pm$ 671  km s$^{-1}$), respectively. For limb events, the mean initial acceleration of RL CMEs for SC 23 (1.05 $\pm$ 0.84 km s$^{-2}$) and SC 24 (1.20 $\pm$ 0.89 km s$^{-2}$) is insignificant. For limb events alone, it is found that the mean initial speeds of RL CMEs for SC 23 and SC 24 are 1212 $\pm$ 500 km s$^{-1}$ and 1460 $\pm$ 904 km s$^{-1}$, respectively. It is also noticed that the final speed of the RL CMEs for SC 23 (1300 $\pm$ 476 km s$^{-1}$) is lower than SC 24 (1575 $\pm$ 1016 km s$^{-1}$). However, the differences in mean values are statistically insignificant. It exactly replicates the RL CME's initial acceleration is the same for both sets of events. The propagation of the LE of the RL CME is inferred to exhibit virtually same behavior in the corona.
\begin{figure} [ht]   %%%%%%%%%%%%%%%%%% FIGURE 2 
\centerline{\includegraphics[width=1.\textwidth,clip=]{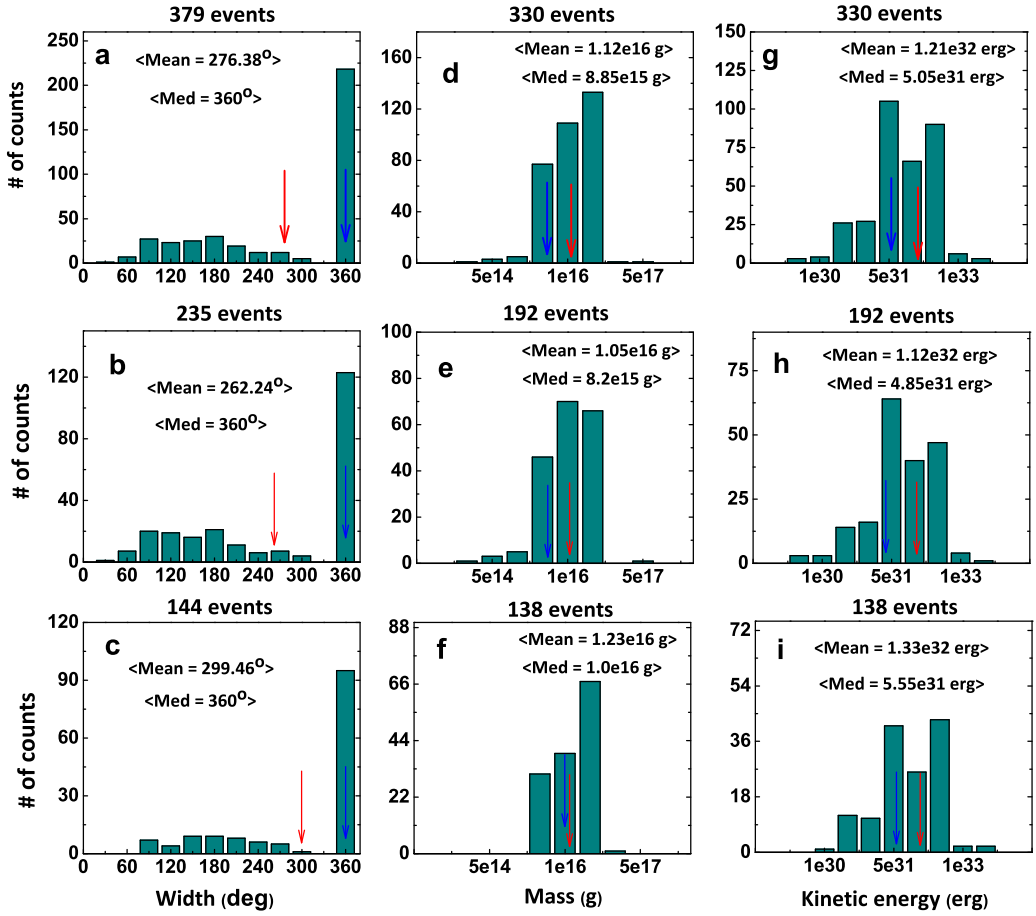}}
\caption{Distributions of (a-c) width, (d-f) mass, and (g-i) kinetic energy for all RL CMEs (top row), SC 23(middle row) and SC 24(bottom row). The mean and median values are marked by red and blue arrows, respectively.}
\end{figure}

The apparent breadth of CMEs appears to be a significant parameter in predicting the kinetic energy of CMEs in general (Gopalswamy et al., 2002). CMEs that were faster and wider than average were also closely connected with DH type II radio emissions (Gopalswamy et al., 2001). CME widths are a major factor influencing space weather events (Kahler et al., 1984). As a result, we analyze the RL CME widths for SC 23 and SC 24. However, determining the true width of halo CMEs is difficult. Because of the projection effect, the true width of halo CMEs cannot be determined from 2D images, at least for the disc event. The disc center was the source of a large proportion of RL CMEs in our analysis. It is impossible to avoid errors when estimating widths, masses, and kinetic energy. Figure 2 shows the distributions of (a-c) apparent widths, (d-f) apparent mass, and (g-i) apparent kinetic energy. As shown in Figure 2a, the mean width of all RL CMEs is 276$^{o}$ $\pm$106$^{o}$, with wide ranges varying from 19$^{o}$ to 360$^{o}$. Furthermore, full halo and partial halo ($>$ 120$^{o}$) account for 58 percent and 30\% of all RL CMEs, respectively, whereas the remaining 12\% of RL CMEs are narrower ($<$120$^{o}$). It is worth noting that the average apparent widths of the RL CMEs in SC 23 (262$^{o}$ $\pm$ 112$^{o}$) are smaller than those in SC 24 (299$^{o}$ $\pm$ 91$^{o}$). The mean difference between SC 23 and SC 24 is statistically significant (P $<$ 0.01\%), according to a student \emph{t}-test. As mentioned earlier, the projection impact is much reduced for limb events. When only limb events are taken into account, the difference in mean widths of RL CMEs for SC 23 and SC 24 (227$^{o}$ $\pm$ 107$^{o}$ and  280$^{o}$ $\pm$ 91$^{o}$, respectively) has a strong statistical significance (P = 1\%). The width of the overall CMEs agrees with this tendency. The mean width of general CMEs during SC 23 is narrower than during SC 24, according to Michalek, Gopalswamy, and Yashiro (2019). It is also confirmed that RL CMEs are expanded further due to the weak state of the heliosphere, even if the velocity of the CMEs are identical in both SC 23 and SC 24.\\[5pt]
Figure 2 (second column) shows the distributions of mass of all RL CMEs (d), SC 23 (e), and SC 24 (f). We found that the mean apparent mass of all RL CMEs is 1.12 $\times$10$^{16}$ $\pm$ 1.18 $\times$ 10$^{16}$g, four times larger than the general CMEs (2.5 $\times$10$^{15}$ g), estimated by Michalek, Gopalswamy, and Yashiro (2019). It provides a wide range of masses, ranging from 6.4 $\times$10$^{13}$ g to 1.6 $\times$10$^{17}$ g. Lamy et al. (2019) found that the mean apparent mass of CMEs over both solar cycles is 1.2$\times$10$^{17}$ g, nearly one order of magnitude smaller than the mean apparent mass of all RL CMEs. However, mean apparent mass of the RL CMEs for SC 23 is almost equal to SC 24. Figure 2g shows the kinetic energy of the all RL CME for both cycles, which exhibits a similar pattern. RL CMEs have an apparent kinetic energy range of 1.3 $\times$ 10$^{29}$ erg to 4.2 $\times$ 10$^{33}$ erg with a mean value of 1.21 $\times$ 10$^{32}$ $\pm$ 2.96 $\times$ 10$^{32}$ erg.  This average value is higher than the kinetic energy of general CMEs observed by the LASCO over these two cycles (5.8 $\times$ 10$^{30}$ erg) which is shown by Lamy et al. (2019). In SC 23 and SC 24, the mean difference in apparent kinetic energy of the RL CMEs is 1.12 $\times$ 10$^{32}$  $\pm$  3.24 $\times$ 10$^{32}$ erg and 1.33 $\times$ 10$^{32}$ $\pm$ 2.53 $\times$ 10$^{32}$ erg, respectively. Lamy et al. (2019) also showed that the mean apparent kinetic energy of general LASCO CMEs in SC 23 (7.6 $\times$ 10$^{30}$ erg) is higher than that of CMEs in SC 24 (4.5 $\times$ 10$^{30}$ erg). We also found that there are no significant differences in the mass and kinetic energy of RL CMEs between SC 23 and SC 24 for limb events. Gopalswamy et al. (2015b) also found that the halo CME kinematics during the rising phase of SC 23 and 24 would not change significantly. This research shows how the reduced total pressure in the heliosphere causes SC 24 RL CMEs to expand abnormally, causing them to decelerate and reach their maximum speed at lower heights than SC 23.Our results for different set of CMEs are consistent with earlier research (Gopalswamy et al., 2015b; Gopalswamy et al., 2020, Anitha, Michalek, and Yashiro, 2020).

\subsection{ Properties of DH type II bursts for  SC 23 and SC 24}

Figure 3 shows the distributions of (a-c) starting frequency, (d-f) end frequency, and (g-i) CME nose height at the time of DH type II starts for all RL CMEs (top row), SC 23 (middle row), and SC 24 (bottom row) events. The starting frequency of the DH type II radio emission is a special feature for determining the height of the shock and giving a potential hint for forecasting space weather events. However, the starting frequencies of most DH type II bursts are near the WAVES instrument's highest cut-off frequency (14 MHz) onboard. Because the type II catalogue was utilized, the STEREO/WAVES data after 2008 has an upper cut-off frequency of 16 kHz, whereas Wind/WAVES data has 14 kHz. As a whole, the upper cut-off starting frequencies of DH type II bursts that are more than 14 kHz have been modified to 14 kHz. It is also shown that the variations in mean starting frequencies of DH type II bursts for SC 23 and SC 24 (9945$\pm$ 4710 kHz and 11670 $\pm$ 4327 kHz, respectively) are still statistically highly significant (P $<$ 0.0001\%). For limb events, the mean difference in DH type II burst starting frequencies for SC 23 (10164 $\pm$ 4777 kHz) is considerably smaller than for SC 24 (12560 $\pm$ 4961 kHz). This difference in mean starting frequencies is also statistically significant (P = 1\%). The higher starting frequency of DH type II bursts for SC 24 is thought to contribute to shock formation at lower heights than for SC 23. Many authors have pointed out that the end frequency of DH type II bursts reflects the energy of CME (Gopalswamy et al. 2005, Prakash et al. 2014, 2017). The distributions of end frequencies for  all RL CMEs associated events (d), SC 23 (e), and  SC 24 (f) are shown in Figure 3 (third column). The end frequencies of DH type II bursts for both solar cycles, as shown in this graph, range from 20 kHz to 12000 kHz, with a mean value of2151 $\pm$ 2741 kHz. It is worth noting that 50\% of DH type II radio bursts occur in the kilometric domain (below 1000 kHz), with the remaining 50\% occurring in the 2000 -- 12000 kHz frequency range. However, we found that the variations in the mean value of end frequencies for SC 23 and SC 24 of DH type II bursts (2363 $\pm$ 2741  kHz and 1809 $\pm$ 2716 kHz, respectively) are statistically significant (P = 5 \%).

\begin{figure} [ht]   %%%%%%%%%%%%%%%%%% FIGURE 2 
\centerline{\includegraphics[width=1.\textwidth,clip=]{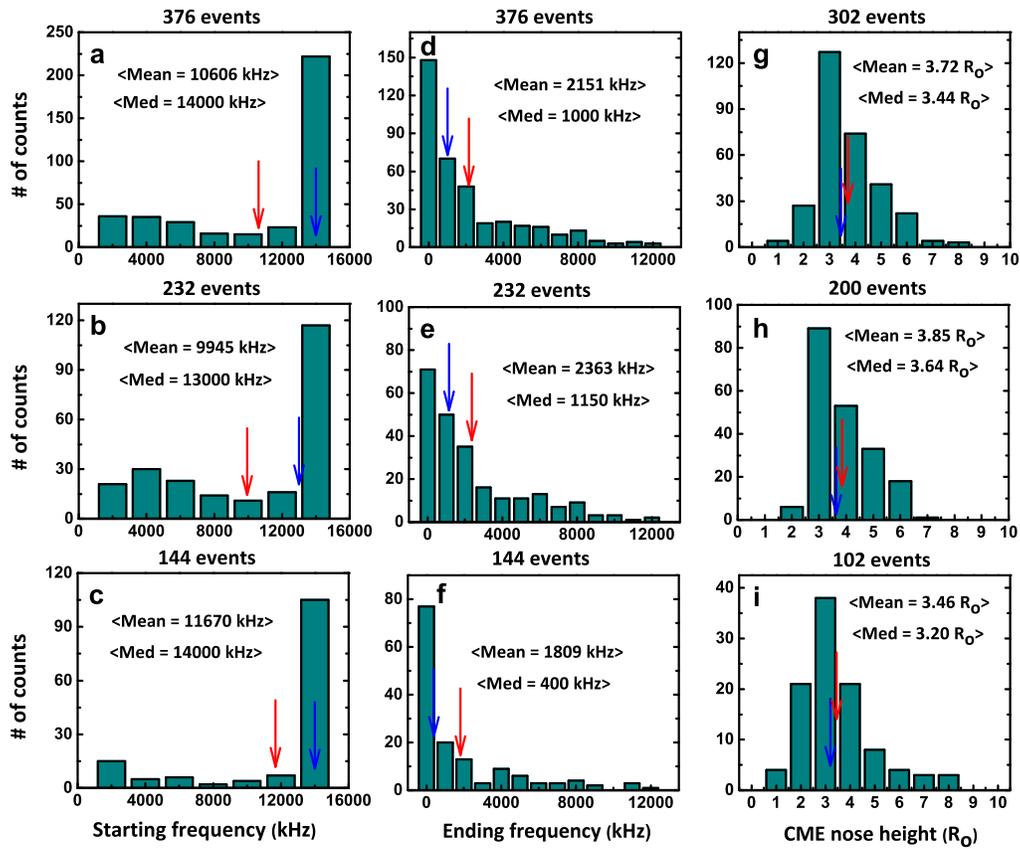}}
\caption{Distributions of (a-c) starting frequency, (d-f) ending frequency, and (g-i) CME nose height at the starting time of DH type II for all RL CME-associated events (top row), SC 23(middle row) and SC 24 (bottom row).The mean and median values are marked by red and blue arrows, respectively.}
\end{figure}

The shock ahead of the CME nose has long been thought to be a better location for DH type II emission (Martinez Oliveros et al., 2012; Shen et al., 2012). Based on the assumption of constant CME speed, we calculated the RL CME nose height at the first observation time of the DH type II emission using back extrapolation of h-t measurements. The shock could have been formed earlier, but the associated type II emission cannot be detected because the emission frequency is above the frequency range of the receiver. For predicting the CME nose height at the time of DH type II bursts, lying beyond the 3 \emph{R}$_{\circledcirc}$ is a preferable assumption. For all RL CMEs (g), SC 23 (h), and SC 24 (i), Figure 3 (third column) depicts the distributions of CME nose heights at the start time of DH type II bursts. The h-t measurements of the leading edge of associated RL CMEs are used to calculate the CME nose height. The CME nose height at the time of DH shock formation was found to be in between 1.3 \emph{R}$_{\circledcirc}$ and 16.39 \emph{R}$_{\circledcirc}$ (mean = 4.07 $\pm$ 2.10 \emph{R}$_{\circledcirc}$). The difference in mean values of RL CMEs nose height between the SC 23 and SC 24 (3.94 $\pm$ 1.32 \emph{R}$_{\circledcirc}$ and 4.28 $\pm$ 2.96 \emph{R}$_{\circledcirc}$, respectively) is statistically insignificant (P = 10\%). It is observed that almost 77\% and 59\% of DH type II bursts associated with RL CME for SC 23 and SC 24 are formed beyond 3 \emph{R}$_{\circledcirc}$. The difference in mean values of RL CMEs nose height for SC 23 and SC 24 are 3.85 $\pm$1.04 \emph{R}$_{\circledcirc}$ and 3.46 $\pm$ 1.48 \emph{R}$_{\circledcirc}$, respectively. This mean difference is statistically significant (P = 5\%). It is inferred from this result that the shock formation height of RL CMEs for SC 24 (more decelerated) is slightly lower than that of SC 23 (less decelerated).

\begin{figure} [ht]   %%%%%%%%%%%%%%%%%% FIGURE 2 
\centerline{\includegraphics[width=1.\textwidth,clip=]{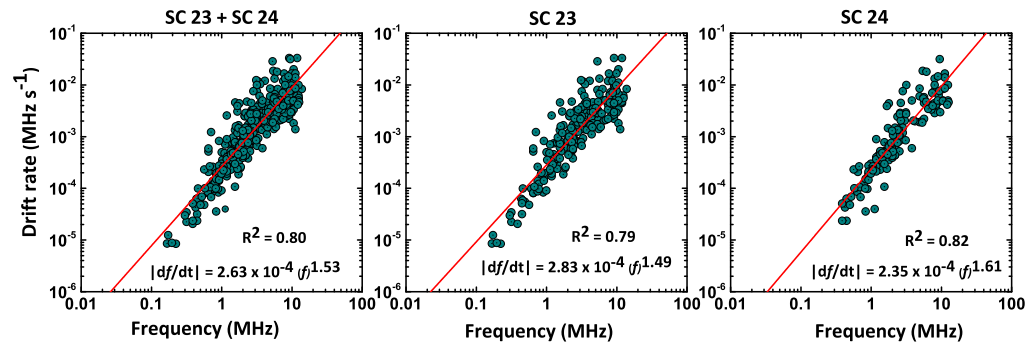}}
\caption{Correlation between the geometric mean frequencies and drift rates of DH type II radio bursts for all RL CME-associated events (left), SC 23 (middle), and SC 24 (right). The power-law is used to fit the data points in the form of $|df/dt|$= Af$^{\epsilon}$.}
\end{figure}

The frequency dependency of the drift rates for DH type II radio bursts was also analysed. In Figure 4, it is present for all RL CMEs-associated DH type IIs (left), SC 23 (middle), and SC 24 (right). The starting and end frequencies, as well as the duration of a DH type II burst are used to estimate the drift rate. It is important to note that the drift rate calculated in this method corresponds to the slope of the type II trajectory secant line at a mid-frequency (fm), not the slope of the tangent line at the starting frequency. As a result, we estimated the mid-frequency as the geometric mean of the starting and end frequencies  of each DH type II burst. We found a distinct relationship between the drift rates and the mid-frequencies of DH type II radio bursts for both solar cycles as shown in Figure 4 (left), with a correlation coefficient of \emph{R} = 0.80 ($\epsilon$ = 1.53). However, as seen from the Figure 4 (right) and 4 (middle), the power -- law indices for SC 24 ($\epsilon$ = 1.61) is significantly higher than for SC 23 ($\epsilon$ = 1.49) with the correlation coefficients \emph{R} = 0.82 and \emph{R} = 0.79, respectively. These correlation coefficients are all in consistent with recent results (Umuhire et al., 2020; Pappa Kalaivani et al., 2021) for a different sets of events. It shows that the close relationship between $df/dt$ and \emph{f} can be caused by the fact that the shock travels at a speed V, emitting at successively lower frequencies determined by the local plasma density (n), which decreases with heliocentric distance (r) as $r^{-2}: |df/dt| = V (df/dr) =V (f/2n)(dn/dr) = Vf^{2}$. The type II emission is assumed to occur at the fundamental plasma frequency (\emph{f} $\sim$ n$^{1/2}$) as explained by Vrsnak et al. (2001). However, there is a clear trend that bursts with higher frequencies have higher drift rates (Aguilar-Rodriguez et al., 2005; Gopalswamy et al., 2009b).

 \subsection{Properties of solar flares associated with  radio-loud CME during SC 23 and SC 24}

The solar source locations of the associated eruptions were reported in the online type II catalogue. It is specified by heliographic coordinates (latitude and longitude) which were taken from the online type II catalogue for this investigation. For example, N20W45 indicates the source location, where N20 denotes northern latitudes of 20$^{o}$ and W45 denotes western longitudes of 45$^{o}$. Figure 5 shows the distributions of solar flare properties: (a-c) latitude, (d-f) longitude, (g-i) peak flux, and (j-l) integrated flux for all RL CMEs associated events (top row), SC 23 (bottom row), and SC 24 (bottom row). For both solar cycles, the corresponding latitudes of RL CMEs are bound to a range of 60$^{o}$. The mean latitude value is --0.70$^{o}$ $\pm$16.78$^{o}$. However, Figure 5a shows that 71\% of all RL CMEs occurred around the equator ($\pm$20$^{o}$). This result is consistent with the streamer belt's spatial distribution. For RL CMEs associated events, the difference in mean latitude for SC 23 and SC 24 is comparable (--0.89$^{o}$ $\pm$ 17.30$^{o}$ and --0.40$^{o}$ $\pm$ 15.98$^{o}$, respectively). However, Michalek, Gopalswamy, and Yashiro (2019) found that the mean latitude of general population of CMEs for SC 23 and SC 24 are --1.8$^{o}$ and 4.5$^{o}$, respectively, contradicting our findings. The majority of the RL CMEs associated events in SC 24 are generated on the solar disk's southern side, according to our results. Figure 5 (second column) shows the longitude (L) distributions for all RL CME-related occurrences (d), SC 23 (e), and SC 24 (f).  The observation that all RL CMEs during both SCs are widely spread over longitudes (($\pm$ 90$^{o}$) is evident from this figure. We found that 40\% of RL CME-related flares originate from the western side ($>$W30$^{o}$), 31\% from the disc center (E30$^{o}$ $\le$ L $\le$ W30$^{o}$), and the remaining 29\% originate from the eastern side ($>$E30$^{o}$).The difference in source longitude mean values for SC 23 and SC 24 (7.29$^{o}$ $\pm$ 59.18$^{o}$ and 4.29$^{o}$ $\pm$ 50.37$^{o}$, respectively) is statistically negligible.

\begin{figure} [ht]   %%%%%%%%%%%%%%%%%% FIGURE 2 
\centerline{\includegraphics[width=1.\textwidth,clip=]{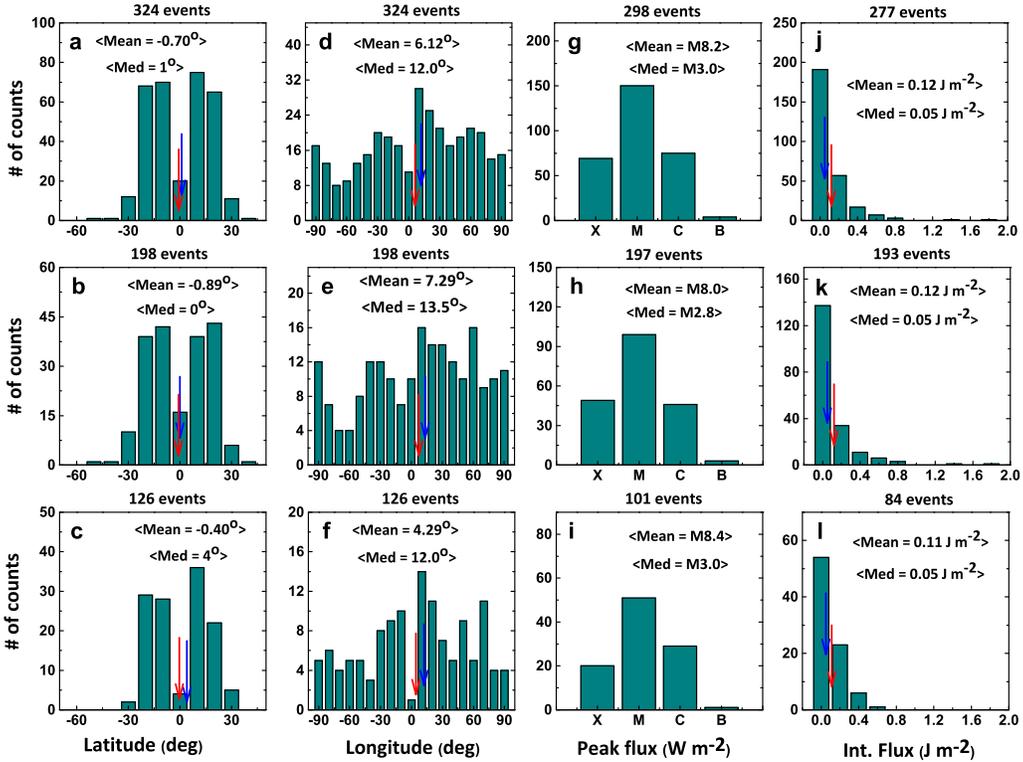}}
\caption{Distributions of (a-c) latitude, (d-f) longitude, (g-i) peak flux, and (j-l) integrated flux of solar flares for all RL CMEs associated events (top row), SC 23 (middle row) and SC 24 (bottom row). The mean and median values are marked by red and blue arrows, respectively.}
\end{figure}

The peak flux distributions of all RL CME-associated flares (g), SC 23 (h), and SC 24 (i) are shown in Figure 5 (third column). It is worth noting that X and M class events account for 23\% and 50\% of the RL CME-related flares, respectively. The mean peak flux of RL CMEs linked flares for SC 23 and SC 24 is 8.01 $\times$ 10$^{-5}$ $\pm$ 1.34 $\times$ 10$^{-4}$ W m$^{-2}$ and 8.41 $\times$ 10$^{-5}$ $\pm$ 1.59 $\times$ 10$^{-4}$ W m$^{-2}$, respectively, statistically insignificant. The differences in mean peak flux of RL CMEs associated flares for SC 23 and SC 24 are statistically negligible, at 8.01$\times$ 10$^{-5}$ $\pm$ 1.34 $\times$ 10$^{-4}$ W m$^{-2}$ and 8.41 $\times$ 10$^{-5}$ $\pm$ 1.59 $\times$ 10$^{-4}$W m$^{-2}$, respectively. Solar flares associated with RL CMEs have a lower peak flux (8.15 $\times$ 10$^{-5}$ W m$^{-2}$) than flares associated with geo-effective storms (0.77 $\times$ 10$^{-5}$ W m$^{-2}$). The integrated intensity of the solar flares associated with all RL CMEs exhibited a similar pattern. SC 23 (0.12 $\pm$ 0.21 J m$^{-2}$) has an integrated flux for solar flares that is almost identical to SC 24 (0.11 $\pm$ 0.13 J m$^{-2}$).The results are consistent with those of Gopalswamy et al. (2015b), who observed no notable change in the size of solar flares associated with halo CMEs. In terms of the speed, mass, and kinetic energy of RL CMEs, as well as peak flux and integrated flux of solar flares, we found that there is nothing distinctive about the RL CMEs in SC 24.

\subsection{The annual variation in the CMEs for SC 23 and SC 24}

\begin{figure} [ht]   %%%%%%%%%%%%%%%%%% FIGURE 2 
\centerline{\includegraphics[width=1.\textwidth,clip=]{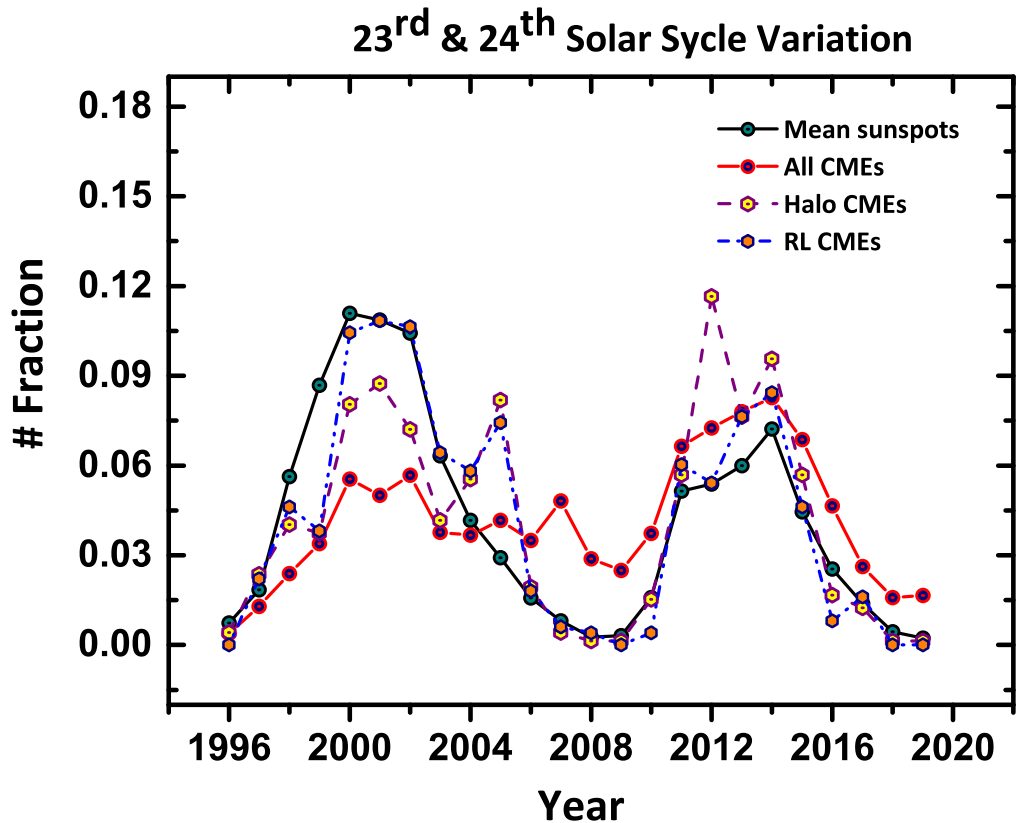}}
\caption{Annual variation of mean sunspot number, number of all general CMEs, halo CMEs, radio loud CMEs for both SC 23 and SC 24}
\end{figure}

Figure 6 presents the annual variation of all, halo, and RL CMEs as well as the mean sunspot number (SSN). In comparison to SC 23, the SSN declined by 31\% in SC 24. Interestingly, the number of all general CMEs in SC 24 is 7\% higher than in SC 23.However, in SC 24, the number of halo and RL CMEs is dropping by 10\% and 30\%, respectively, in the declining phase. From 1997 to 2000, the average number of sunspots increased, then decreased until it reached the solar minimum in 2008. In the year 2000, it reached a maximum of 173.9. SC 24, on the other hand, hits its peak in 2014, with a mean value of 113.3. It should be mentioned that SSN decreased by 21\% at the maximum of SC 24 compared to SC 23.
During the solar maximum in SC 24, the rate of halo CMEs is increased by 9\% whereas the rate of RL CMEs decreased by 11\% compared to SC 23. It is well consistent with both sunspot cycles in the case of all general CMEs. During the period 2007 -- 2009, however, we noticed a small variation. Despite the low number of sunspots, the SOHO/LASCO spacecraft observed a considerable number of CMEs during the solar minimum. However, there are small variations in the halo and RL CMEs for both cycles. We found that there are two peaks for halo and RL CMEs during the solar maximum of SC 23 and SC 24. The disparity in annual variation between the mean sunspot number (decreasing trend) and the number of all CMEs, halo, and RL CMEs during the period 2004 -- 2006 is noteworthy (maximum). Shanmugaraju et al. (2021) found that the occurrence rate of RL CMEs in SC 24 is 0.52 times lower than in SC 23. This result is very identical with what we found.
\section{Summary and conclusions}
A set of 379 RL CMEs is statistically analysed ($\ge$ 3 height-time measurements) along with their related solar flares listed in CDAW's type II catalogue from August 1996 to December 2019, spanning both SC 23 and SC 24. The sample events are classified into two groups: \emph{i)} 235 events in SC 23 (August 1996 -- December 2008), and \emph{ii)} 144 events in SC 24 (January 2009 -- December 2019). This basic features and kinematics of RL CMEs, as well as their associated solar flares and DH type II bursts, are analyzed. The characteristics of these events are compared for SC 23 and SC 24. The key findings of our study can be summarized as follows:
\begin{itemize}
    \item {It is interesting to note that the RL CMEs for SC 24 (--17.39 $\pm$ 43.51 m s$^{-2}$) are more decelerated than those of SC 23 (--8.29 $\pm$ 36.23 m s$^{-2}$).} 
    \item {There is a statistically significant difference in the mean peak speed of the RL CMEs in SC 23 and SC 24 (1443 $\pm$ 504 km s$^{-2}$ and 1920 $\pm$ 649 km s$^{-2}$, respectively)} 
    \item {The RL CMEs in SC 23 (13.82 $\pm$ 7.40 \emph{R}$_{\circledcirc}$) averaged a slightly higher peak speed height than the RL CMEs in SC 24 (12.51 $\pm$ 7.41 \emph{R}$_{\circledcirc}$).}
    \item {There are no significant differences in the sky-plane and space speeds of the RL CMEs for SC 23 and SC 24 limb events.}
    \item {For both SC, 58\% RL CMEs are full halo and 30\% of RL CMEs are partial halo.  It is worth noting that the mean apparent widths of the RL CMEs in SC 23 (262$^{o}$ $\pm$ 112$^{o}$) are relatively lower than those in SC 24 (299$^{o}$ $\pm$ 91$^{o}$), which is statistically significant.}
    \item{The mean apparent mass of RL CMEs in SC 23 is, however, slightly lower than that of RL CMEs in SC 24.} 
    \item{It is found that differences between the mean starting frequencies of DH type II bursts for SC 23 and SC 24 (9944 $\pm$ 4710 kHz and 11670 $\pm$ 4327 kHz, respectively) is statistically significant (P $<$ 0.0001\%).} 
    \item {It is also found that the difference in the mean ending frequencies (2363 $\pm$ 2741  kHz and 1809 $\pm$ 2716 kHz, respectively) for SC 23 and SC 24 of DH type II burst is also statistically significant (P = 5\%). SC 23 has a lower average CME nose height at the start of DH type II bursts than SC 24. The difference in the mean nose height of the CMEs for SC 23 and SC 24 is statistically significant (3.85 \emph{R}$_{\circledcirc}$ and 3.46 \emph{R}$_{\circledcirc}$, respectively).}
    \item {For the total sample of both solar cycles, there is a significant relation between the drift rates and the mid-frequencies of DH type II radio bursts, with a correlation coefficient of  0.80 ($\epsilon$ = 1.53). SC 24 has a significantly higher power law index value $\epsilon$ = 1.61 than SC 23 ($\epsilon$ = 1.49).} 
    \item {In comparison to SC 23, the SSN declined by 47\% in SC 24.  But, when compared to SC 23, the frequency of occurrence of general CMEs in SC 24 has increased by 7\%. However, in SC 24, the quantity of halo and RL CMEs tend to decrease by 10\% and 30\%, respectively.} 
    \item {This study revealed the effect of the reduced total pressure in the state of the heliosphere for SC 24 which allows RL CMEs to expand wider and decelerate more and reach their peak speed at lower height than SC 23. It also shows that the DH type II radio emissions originate of SC 24 at higher starting frequencies than those of SC 23.}
    \end{itemize}

%%%%%%%%%%%%%%%%%%%%%%%%%%%%%%%%%%%%%%%%%%%%%%%%%%%%%%%%%%%%%%%%%%%%%%%%%%%
\begin{acks}
We thank the reviewer for giving constructive comments to improve the quality of this manuscript. We greatly acknowledge the data support provided by various online data centers of NOAA and NASA. We would like to express our gratitude for the CDAW's Wind/WAVES type II catalogues. The SOHO/LASCO CME catalog is generated and maintained at the CDAW Data Center by NASA and The Catholic University of America in cooperation with the Naval Research Laboratory. SOHO is a project of international cooperation between ESA and NASA.
 \end{acks}

%and similarly for ending page numbers.
%

\end{article} 

\end{document}